\documentclass{article}
\usepackage{nicefrac}
\usepackage{amsfonts}
\usepackage{amssymb}
\usepackage{amsmath,mathrsfs,dsfont}
\usepackage{graphicx}
\usepackage{appendix}
\usepackage{cite}
\usepackage{color}
\newcommand{\D}{D\!\!\!\!/}

\begin{document}

\title{On Noncommutative Effects in Entropic Gravity}

\author{C.~M.~Gregory$^a$\thanks{carogregory@gmail.com} and A.~Pinzul$^{a,b}$\thanks{apinzul@unb.br}\\
\\
$^a$ \emph{Instituto de F\'{\i}sica, Universidade de Bras\'{\i}lia}\\
\emph{70910-900, Bras\'{\i}lia, DF, Brasil}\\
\\
$^b$ \emph{{International Center of Condensed Matter Physics} }\\
\emph{C.P. 04667, Brasilia, DF, Brazil} \\
}
\date{}
\maketitle

\begin{abstract}
We analyze the question of possible quantum corrections in the entropic scenario of emergent gravity. Using a fuzzy sphere as a natural quasiclassical approximation for the spherical holographic screen, we analyze whether it is possible to observe such corrections to Newton's law in principle. The main outcome of our analysis is that without the complete knowledge of quantum dynamics of the microscopical degrees of freedom, any Plank scale correction cannot be trusted. Some perturbative corrections might produce reliable predictions well below the Plank scale.
\end{abstract}

\section{Introduction}

The final form of Quantum Gravity (QG) is yet to be found. Either of the two main candidates for such a theory, (Super)String Theory and Loop Quantum Gravity (LQG), despite much progress, still cannot be taken as the final answer (the very existence of two seriously different theories of QG means that the problem of the quantization of gravity is far from being settled). In this situation, any effort in this direction should be welcomed. In particular, given a model for QG, it is very important to understand how the classical limit of Einstein's General Relativity (GR) emerges, as well as to learn how to calculate possible quantum-gravitational corrections. Concerning the classical limit, the recently proposed entropic gravity \cite{Verlinde:2010hp} might turn out to be quite important. In this model, E.~Verlinde proposes that gravity, instead of being a fundamental force, has an emergent, entropic origin. Roughly speaking, one can think of gravitational force as being caused by the change of entropy of the system - holographic screen plus a test mass. Under some assumptions, this model uniquely leads to the classical GR for quite general quantum dynamics. E.g., in \cite{Smolin:2010kk}, it was shown how using this approach, one can get Newton's law in the framework of LQG. Due to this universality, the question of quantum-gravitational corrections becomes very important. The hope is that this will eventually allow us to tell between different models of QG. In this regard, the following observation is of great importance \cite{Doplicher:1994tu}: independently of specific details of the final theory of QG, the quasiclassical regime, i.e. when typical energy scale is close to the Plank scale, should be described by a field theory on some noncommutative space-time. This should capture some nonperturbative effects of QG. The details of this noncommutativity do depend on the QG model. In 3d, this was explicitly demonstrated in \cite{Freidel:2005bb} for Ponzano-Regge model.

The main goal of our work is to analyze possible effects of the underlying noncommutativity on the entropic gravity. Ever since its publication, entropic scenario generated a series of comments and criticisms. One of the main objection is that the obtained results are a consequence more of the dimensional analysis rather then some fundamental physical reasons (this especially applies to Newton's law). This makes it very important to analyze the process of the ``interaction'' of a test particle with the holographic screen, i.e. how this particle becomes a part of the screen. From our point of view, here one has one of the major problems of the model: a well defined smooth holographic screen, e.g. a sphere, is an adequate model to reproduce the GR limit and is too restrictive if one wants to go beyond the classical approximation. Taking into account the above observation about the universality of noncommutativity in QG, we address this point by considering a fuzzy sphere as a natural candidate for a holographic screen, which ``remembers'' about its quantum-gravitational origin.

As the detailed answer to the above question on how a test particle becomes a part of the screen depends on our knowledge of quantum dynamics of the microscopic degrees of freedom, we will be interested in a less ambitious problem: how does a test particle ``see'' the holographic screen (which is taken to be a fuzzy sphere)? As the main tool of our analysis, we use some methods of spectral geometry. Our main results are following:

1) To actually discuss quantum corrections close to the Plank scale, one needs to know the details of quantum dynamics. Any attempt to obtain such corrections without knowing how the test particle ``sees'' the holographic screen will be destroyed by the uncertainties due to our ignorance about this process. On the other hand, away from the Plank scale, some universal corrections might be well-defined, confirming their perturbative origin.

2) Somewhat related and more important point: now, a test particle is a very important ingredient of the whole construction. One cannot remove it from the picture in principle. We will see that there is regime (though quite beyond any experimental reach) when different test particles will see the holographic screen quite differently. Hence, one can speculate about a possible violation of the equivalence principle by quantum gravity (in this scenario).

The plan of the paper is following: In section \ref{IntroVerlinde} we briefly discuss Verlinde's approach to entropic gravity, its consistency and some possible corrections. After arguing that to go beyond the classical limit one has to abandon the notion of a smooth holographic screen in favor of a noncommutative one, in section \ref{Fuzzy} we introduce fuzzy sphere and review some properties of its Dirac operator.  In section \ref{WeylSec}, we apply (the generalization of) Weyl's theorem to calculate the area of a fuzzy holographic screen as seen by a test particle. We analyze possible corrections to the classical area in the regimes of weak and full noncommutativity. Section \ref{Discussion} contains the discussion and interpretation of the obtained results. We conclude with the summary and some final remarks.

\section{Classical Entropic Gravity and Beyond}\label{IntroVerlinde}

Here we briefly summarize the main steps and inputs leading to the entropic scenario \cite{Verlinde:2010hp}. We stress on the points that need, from our point of view, more justification or more careful analysis.

1) One starts by assigning to any surface some entropy, which scales as area, $A$. In such a way, any surface, not just black hole horizons, plays a role of a ``holographic screen''.

2) When a test particle (which is assumed to be elementary, i.e. pointlike) of mass $m$ approaches such a holographic screen by distance of the order of the Compton wavelength, $\lambda_m = \frac{\hbar}{mc}$, the entropy of the screen is increased by $\Delta S = 2\pi k_B$. To simulate more continuous change in entropy, one assumes that when the particle is at distance $\Delta x$ from the screen, the change is given by
\begin{eqnarray}\label{entropychange}
\Delta S = 2\pi k_B \frac{\Delta x}{\lambda_m}\ .
\end{eqnarray}
One immediately sees tension between the first two assumptions \cite{Modesto:2010rm}: while in 1) entropy scales like area, according to (\ref{entropychange}) the scaling is as distance. Already this suggests that one should have a better understanding of how exactly a test particle becomes a part of the holographic screen.

3) The energy associated to the screen, $E_S$, is given by the total energy inside the screen. In non-relativistic limit, $E_S = Mc^2$, where $M$ is the total mass encircled by the surface.

4) This energy is equally distributed between $N$ quanta of the surface, which leads to the temperature of the screen:
\begin{eqnarray}
E_S = \frac{1}{2}N k_B T\ ,
\end{eqnarray}
where $N=\frac{A}{l^2_P}$ and $l_P = \sqrt{\frac{G\hbar}{c^3}}$ is the Plank length.

5) The last assumption is that the resulting entropic force
\begin{eqnarray}
F = T\frac{\Delta S}{\Delta x}
\end{eqnarray}
{\it is} gravity.

Combining 1)-5) one immediately arrives (for a spherically symmetric configuration) at Newton's law \cite{Verlinde:2010hp}. One of the most attractive features of this scenario is its universality: independently of the actual microscopic dynamics, as long as the fundamental theory satisfies 1)-5), it will lead to General Relativity. This has already been used as a possible way to get Newton's law from LQG \cite{Smolin:2010kk}.

We have already mentioned one potential tension between some of the assumptions 1)-5). Another one was raised by many critiques of the entropic scenario: with this setup, Newton's law is a mere consequence of the dimensional analysis. This makes it crucial to check the emergence of gravity in this way from some fundamental theory or at least to go beyond the classical limit and try to calculate quantum corrections. The major effort in this direction has been based on calculating corrections to the entropy and then using this quantum corrected entropy in derivation of corrected Newton's law. Here we mention just a couple of works, which are relevant for our consideration.

In \cite{Modesto:2010rm}, LQG-inspired corrections were considered in the form
\begin{eqnarray}\label{LQGentropy}
S = \frac{A k_B}{4l^2_P} - a k_B\ln\left(\frac{A}{l^2_P}\right) + b k_B\left(\frac{A}{l^2_P}\right)^{3/2} \ ,
\end{eqnarray}
where $a$ and $b$ are some constants of order of one. While the first term in (\ref{LQGentropy}) is the usual Bekenstein entropy, the others represent corrections. The logarithmic correction is quite universal, while the volume correction is motivated by LQG. The use of (\ref{LQGentropy}) leads to the following corrections to Newton's law:
\begin{eqnarray}\label{Newtoncorrected1}
F = -\frac{GMm}{R^2}\left( 1 - a \frac{l^2_P}{\pi R^2} + 12 b \sqrt\pi \frac{R}{l_P}\right) \ .
\end{eqnarray}
Due to the universality of the logarithmic correction in (\ref{LQGentropy}), the first corrction to Newton's law is also quite universal and follows from different models for quantum corrections \cite{Donoghue:1994dn,Hamber:1995cq,Floratos:1999bv}. The volume correction is interesting  because it leads to much stronger gravity at large distances, which has potential of explaining anomalous galactic rotational curves \cite{Modesto:2010rm} (though quick inspection shows that this is the case for quite unnatural values of the parameters $a$ and $b$).

The approach taken in \cite{Nicolini:2010nb} uses some specific model to calculate corrections to entropy due to noncommutativity of space-time. One arrives at the corrected Newton's law
\begin{eqnarray}\label{Newtoncorrected2}
F = -\frac{GMm}{R^2}\left( 1 + \frac{R e^{-R^2/(4\alpha l_P^2)}}{\sqrt{\pi\alpha}l_P}+\frac{R^2e^{-R^2/(2\alpha l_P^2)}}{2\alpha\sqrt{\pi}l^2_P}\right) \ .
\end{eqnarray}
Note that in this case the corrections are exponentially suppressed by $\frac{R^2}{l^2_P}$. This means that even the hypothetical possibility to measure such corrections becomes even more evasive.

We mention these two works, because, from our point of view, they are trying to address two very important issues of the original entropic scenario: while \cite{Modesto:2010rm} has something to say about the process of a test particle becoming a part of the holographic screen, \cite{Nicolini:2010nb} considers noncommutative space-time, which is more natural from the quantum gravity point of view \cite{Doplicher:1994tu}.

In this paper, we also take this ``noncommutative'' point of view: we model our spherical holographic screen by a fuzzy sphere $S_F$ \cite{Madore:1991bw} in place of a smooth $S_2$. But instead of considering corrections to entropy (which anyway is beyond our reach without knowing something about quantum degrees of freedom), we ask the following question: how does a test particle see this sphere? In particular, we will try to analyse what is the area of the holographic screen as seen by a test particle. It should be clear that this is closely related to the question of how a test particle ``interacts'' with the screen, though the complete answer to this question once again requires the information about microscopic dynamics.

Before we discuss some properties of fuzzy sphere and its Dirac operator, we would like to give some arguments on why we think a fuzzy sphere should be a natural candidate for a spherical screen as well as why we need its Dirac operator.

1) As we have already stressed several times, the prediction of noncommutativity of the space-time at some quasi-classical regime is model independent \cite{Doplicher:1994tu}.

2) For a spherically symmetric classical distribution of a mass $M$, we should expect that some notion of spherical symmetry is left even in this quasi-classical regime. Recalling that fuzzy sphere is essentially a unique deformation of commutative one such that carries the usual, un-deformed, action of $SO(3)$, we conclude that if a spherical holographic screen should be deformed, a fuzzy sphere is the most natural candidate for this.

3) A fuzzy sphere has a natural, built-in, discreetness, which should be compared with the assumed discreetness of a holographic screen.

4) Using the same kind of arguments, it was speculated, see e.g. \cite{Dolan:2004xd}, that a fuzzy sphere can be quite useful and natural in black hole physics.

5) Why Dirac operator? Below we will discuss this in more details and here we just give some motivation. In our previous work \cite{Pinzul:2010ct}, we showed the effectiveness of a physically relevant Dirac operator for calculations of such geometrical characteristics as area and dimension in the case of deformed geometries (on the example of Horava-Lifshitz deformation of GR). Here we would like to adopt the same procedure for the calculation of ``physical'' area of a fuzzy sphere as seen by a test particle.

\section{Fuzzy Sphere and its Dirac Operator}\label{Fuzzy}

Fuzzy sphere \cite{Madore:1991bw} provides a very important example of a noncommutative space, such that the commutative isometry, $SO(3)$, remains to be symmetry on the noncommutative level. The algebra of the $(N+1)\times (N+1)$ fuzzy sphere of the radius $R$ is generated by noncommutative coordinates:
\begin{eqnarray}
[\mathbf{x_i},\mathbf{x_j}]=i \lambda \epsilon_{ijk}\mathbf{x_k},
\end{eqnarray}
where $\lambda = \frac{2R}{\sqrt{N(N+2)}}$. The presence of two parameters, $R$ and $N$, allows to recover as special limits not only a commutative sphere, but also the Moyal and commutative planes \cite{Alexanian:2000uz}. It is clear from the definition that the noncommutative coordinates, re-scaled by $\lambda$, are given by $(N+1)$-dimensional irreducible representation of the $SU(2)$ generators, $\mathbf{L_i}$.

To proceed with our goal, we will need a Dirac operator. This is one of two of the most important operators in noncommutative geometry (the second one being chirality operator). It is essential for the construction of the differential and integral calculus. As such, it is subject to several natural conditions (see \cite{Connes:1994yd} on the role of the Dirac operator in noncommutative geometry). In the case of fuzzy sphere, we have one extra condition: because this noncommutative space is rotationally invariant, it would be quite desirable that the corresponding Dirac operator respects this symmetry. There are essentially two slightly different proposals for such an operator \cite{CarowWatamura:1996wg,Grosse:1994ed} (see also \cite{Pinzul:2001fk} for the treatment of a more general case of a $q$-deformed sphere, which reduces to \cite{CarowWatamura:1996wg} in a special limit). In this paper, we will work with the operator defined in \cite{CarowWatamura:1996wg}:\footnote{The other choices should not seriously change our conclusions, see the discussion section.}
\begin{eqnarray}\label{DiracWatamura}
\D = \frac{1}{R}\sqrt{\frac{N(N+2)}{(N+1)^2}}\left( (\Lambda+1)-\frac{1}{2R^2}\{\chi , \mathbf{x_i L_i}\}-\frac{\lambda}{R^2}\mathbf{x_i L_i} \right)\ ,
\end{eqnarray}
where $\Lambda = \mathbf{L_i}\otimes \sigma_i$, $\chi = \mathbf{x_i}\otimes \sigma_i$, $\sigma_i$ are the Pauli matrices and the rest of the operators are understood as, e.g., $\mathbf{x_i}=\mathbf{x_i}\otimes\mathds{1}$ etc. It is clear that (\ref{DiracWatamura}) respects $SU(2)$ symmetry of fuzzy sphere. In addition, it anti-commutes with the natural chirality operator, which is given by a linear function of $\chi$ \cite{CarowWatamura:1996wg}. From the definition (\ref{DiracWatamura}), it should be clear, that the Dirac operator is acting, as in the commutative case, on the space of the 2-component spinors. So, we still have the commutative relation between the dimension of the space and the dimension of the spinor bundle. This will be important in our further discussion. Using the rotational invariance and the mentioned above fact that $\mathbf{x_i}$'s are proportional to $\mathbf{L_i}$'s, after some standard algebra, one arrives at the spectrum of the Dirac operator \cite{CarowWatamura:1996wg}
\begin{eqnarray}\label{spectrum}
 \omega_{j\pm}=\pm\frac{1}{R}\Big(j+\frac{1}{2}\Big)\Bigg\{1-\frac{1}{N(N+2)}\Big[\Big(j+\frac{1}{2}\Big)^2-1\Big]\Bigg\}^{\frac{1}{2}}\ ,
\end{eqnarray}
where $j(j+1)$'s are the eigenvalues of square of the total angular momentum $\mathbf{J_i}^2$, where $\mathbf{J_i}:=\mathbf{L_i}+\frac{1}{2}\sigma_i$. Then, from the fact that we are working with $(N+1)$-dimensional irreducible representation, it easy to see that $j\in\mathds{N}$ or $j\in\mathds{N}+\frac{1}{2}$ depending on whether $N$ is odd or even and $0\le j\le \frac{N+1}{2}$.\label{odd-even}\footnote{In \cite{CarowWatamura:1996wg}, it is also argued that to have a correct commutative limit, we should keep only the case of the even $N$, but this will not be important for our calculations. Moreover, we believe that if a fuzzy sphere should come from some theory of quantum gravity, both representations should be allowed.}

\section{Spectral Area of Fuzzy Sphere}\label{WeylSec}

In this section, we show how the Dirac operator (\ref{DiracWatamura}) with the help of Weyl's theorem can be used to calculate the corrections to the area of a fuzzy sphere. Let us start with the classical formulation of Weyl's theorem for the case of commutative geometry and then we will argue that it still makes sense to use this theorem (but now as the definition) for the analysis of some properties of noncommutative or other generalized geometries.\\
\\
\textbf{Weyl's Theorem}
\textit{Let $\Delta$ be the Laplace operator on a closed Riemannian manifold $\mathcal{M}$ of dimension $\mathrm{n}$. Let $N_\Delta (\omega)$ be the number of eigenvalues of $\Delta$, counting multiplicities, less then $\omega$, i.e. $N_\Delta (\omega)$ is the counting function
\begin{eqnarray}\label{counting}
N_\Delta (\omega) := \#\{ \omega_k (\Delta)\ :\ \omega_k (\Delta)\leq \omega\}\ .
\end{eqnarray}
Then
\begin{eqnarray}\label{Weyl}
\lim_{\omega\rightarrow\infty}\frac{N_\Delta (\omega)}{\omega^{\frac{\mathrm{n}}{2}}}=\frac{Vol (\mathcal{M})}{(4\pi)^{\frac{\mathrm{n}}{2}}\Gamma(\frac{\mathrm{n}}{2}+1)}\ ,
\end{eqnarray}
where $Vol (\mathcal{M})$ is the total volume of the manifold $\mathcal{M}$.}\\
\\
Though the theorem is given in terms of the Laplace operator, with the help of the Lichnerowicz formula, $ \D\ \!^2 = \Delta + \frac{1}{4}\mathcal{R}\ $, $\mathcal{R}$ being curvature, it could be easily re-written in terms of the Dirac operator, $\D$. In this case, one should take care of the dimension of the spinor bundle, which is, in the commutative case, equal to $2^m$, where $\mathrm{n}=2m$ or $\mathrm{n}=2m+1$. So, we can see, that in the commutative case, Weyl's theorem provides a way of the simultaneous calculation of both, the volume and the dimension of a manifold. The advantage of this method is in the fact that it is purely algebraic, which allows immediate generalizations to the cases when the usual geometrical techniques do not exist. Before we proceed with the application to the fuzzy sphere case, we would like to give some justifications of such an application.

At the first sight, the na\"{i}ve application of Weyl's theorem to a geometry given by the finite matrix algebra (as in our case) does not seem as something correct. Nevertheless, we will argue that using this theorem, but now as the {\it definition}, still makes sense even in this case. But now, as we will see, one has to clearly distinguish between the formal mathematical and applied physical approaches.
\begin{itemize}
\item The mathematically meaningful application of Weyl's theorem to finite matrix models seems quite doubtful. This could be understood as follows: when the spectrum is unbounded (in particular $N_\Delta (\omega) \rightarrow \infty$), the requirement that the right hand side of (\ref{Weyl}) makes sense (i.e. finite for compact geometries) fixes uniquely the dimension $\mathrm{n}$, which, in its turn, allows to determine $Vol (\mathcal{M})$. But for the case of a finite model, the spectrum is finite, as in (\ref{spectrum}), and $N_\Delta (\omega)$ is finite too. As the result, we do not have any requirement that could fix neither $\mathrm{n}$ nor $Vol (\mathcal{M})$. Here it is crucial that $\omega$ can (and should) be taken arbitrary large. We will see how the situation changes in the presence of the physically motivated cut-off.
\item Let us now use Weyl's theorem as a physical tool to {\it measure} the dimension and area of some, possibly noncommutative, space. We are going to use for this the experimental spectrum of the corresponding Dirac operator. Clearly, this spectrum could be measured only up to some cut-off, $\omega_{\mathbf{co}}$. Typically, even in the case of a finite model, this cut-off is below the maximal eigenvalue of the Dirac operator. So, the apparatus used to probe geometry will not know whether the spectrum is finite or not. Then we can continue to use Weyl's theorem, but now instead of the mathematical limit, $\omega\rightarrow\infty$, we should take the ``physical'' one, $\omega\leq \omega_{\mathbf{co}}$. (See more discussion below.) Now, in general, both volume and dimension will depend non-trivially on the cut-off and without some further (physical) input it is impossible to determine both of them. If we assume the classical value for the dimension, as we will do in this paper, then we can derive the cut-off dependent corrections to the classical volume. In \cite{Pinzul:2010ct}, this approach was successfully used to analyze UV/IR behavior of the spectral dimension in Ho\v{r}ava-Lifshits models of gravity. (See also \cite{Pinzul:2010ct}, especially concluding section, for the discussion and physical interpretation of this approach.)
\end{itemize}

After these comments, let us apply this approach to the case of fuzzy sphere. First of all, we would like to give two arguments in favor of why we want to keep the dimension of the fuzzy sphere equal to the classical one, $\mathrm{n}=2$. Firstly, as we commented after (\ref{DiracWatamura}), the noncommutative Dirac operator acts in the space of the 2-component spinors. This means that the passage between the formulation of Weyl's theorem in terms of the Laplace operator $\Delta$ and the one in terms of the dirac operator $\D$ is the same as in the case of $\mathrm{n}=2$. Secondly, if we look at this from the physical point of view, then during the process of measuring of the spectrum of $\D$ (or $\Delta$), we are already assuming that we are measuring the spectrum of some operator defined on some $2d$ surface. Then any deviations from the commutative result, we treat as the quantum geometrical corrections to the area. Keeping this in mind, let us proceed.

As the first step, we need to calculate the counting function (\ref{counting}). For this, we need to calculate $j$ as a function of $\omega$. Inverting the equation (\ref{spectrum}), we obtain
\begin{eqnarray}\label{jlambda}
\Big(j+\frac{1}{2}\Big)^2=\frac{(N+1)^2\pm[(N+1)^4-4\omega^2R^2N(N+2)]^{1/2}}{2}\ .
\end{eqnarray}

To choose the correct sign in (\ref{jlambda}), we note that $\big(j+\frac{1}{2}\big)^{2} \leqslant \frac{(N+2)^2}{4}$. This leads to the choice of the minus sign. Then we have the maximal value of $j$ corresponding to a cut-off scale $\omega_{\mathbf{co}}$:
\begin{eqnarray}\label{jmax}
\Big(j_{max}+\frac{1}{2}\Big)^2=\frac{(N+1)^2 - [(N+1)^4-4\omega_{\mathbf{co}}^2R^2N(N+2)]^{1/2}}{2}\ .
\end{eqnarray}
Taking into account that the degeneracy of each eigenvalue is equal to $(2j+1)$, we can write the counting function as
\begin{equation}
 N_{|\D|}(\omega_{\mathbf{co}})= 2\sum^{j_{max}}(2j+1)\ .
\end{equation}
The coefficient of 2 comes from the plus/minus sign in (\ref{spectrum}). Taking, for definiteness, $j$ to be half integer, i.e. $N$ to be even (see the footnote on the page \pageref{odd-even}), we obtain the following exact expression for the counting function:
\begin{eqnarray}\label{counting2}
& & N_{|\D|}(\omega_{\mathbf{co}}) = 2\Big(j_{max}+\frac{1}{2}\Big)^2+2\Big(j_{max}+\frac{1}{2}\Big) \nonumber \\
& = & (N+1)^2\Bigg[1-\Bigg(1-\frac{4\omega_{\mathbf{co}}^2R^2(N+2)N}{(N+1)^4}\Bigg)^\frac{1}{2}\Bigg]
+ \sqrt{2}(N+1)\Bigg[1-\Bigg(1-\frac{4\omega_{\mathbf{co}}^2R^2(N+2)N}{(N+1)^4}\Bigg)^\frac{1}{2}\Bigg]^{\frac{1}{2}} \ .
\end{eqnarray}

\subsection{Commutative Limit, $N \rightarrow \infty$}\label{CommutativeLimit}

Let us first use (\ref{counting2}) to reproduce the commutative result for the area of a sphere. This will later help us clarify some points about the applicability of the method, as well as provide the example of the effectiveness of Weyl's theorem.

The commutative limit corresponds to sending the dimension of the representation, $N$, to infinity, while keeping $\omega_{\mathbf{co}}$ finite (at the end, it will also be sent to infinity or, rather, made ``big enough"). In this limit, we have
\begin{equation}\label{classical}
\Big(j_{max}+\frac{1}{2}\Big)^2= R^2 \omega_{\mathbf{co}}^2 \ .
\end{equation}
So, the counting function (\ref{counting2}) becomes
\begin{equation}\label{countingcommut}
 N_{|\D |}(\omega_{\mathbf{co}})= 2R^2\omega_{\mathbf{co}}^2+2R\omega_{\mathbf{co}} \ .
\end{equation}
Now we would like to use Weyl's theorem (\ref{Weyl}) in the form suited for the Dirac operator, $\D $ (setting $\mathrm{n}=2$)
\begin{eqnarray}\label{WeylDirac}
\lim_{\omega_{\mathbf{co}}\rightarrow\infty}\frac{N_{|\D |}(\omega_{\mathbf{co}})}{\omega_{\mathbf{co}}^2}=\frac{2 Area (\mathcal{M})}{4\pi\Gamma(2)}\ ,
\end{eqnarray}
where the factor of 2 is the dimension of spinors (see the discussion after (\ref{Weyl})). Using (\ref{countingcommut}) and (\ref{WeylDirac}) one immediately obtains the well-known result for the area of a commutative sphere $S_2$
\begin{eqnarray}\label{areacommut}
Area(S_2)=4\pi R^2.
\end{eqnarray}
What happens if we take instead of the exact limit in (\ref{WeylDirac}) just a ``physical" one, i.e. take $\omega_{\mathbf{co}}$ very large but finite? How big should $\omega_{\mathbf{co}}$ be so we still could conclude that the area is given, within experimental uncertainty, by the formula ($\ref{areacommut}$)? From (\ref{countingcommut}), we have
\begin{eqnarray}\label{exact}
\frac{N_{|\D |}(\omega_{\mathbf{co}})}{\omega_{\mathbf{co}}^2}= 2R^2 + \frac{2R}{\omega_{\mathbf{co}}}\ .
\end{eqnarray}
Then if $\omega_{\mathbf{co}}\gg 1/R$, or $j_{max}\gg 1$, the ``physical" area will be exactly given by (\ref{areacommut}). The second term in (\ref{exact}), which is the correction to the commutative answer, is nothing but the physical uncertainty in measuring $R$ using a test particle of mass $m$ as a device. Really, we have: $\Delta R \sim \lambda_m \Rightarrow \Delta S \sim \lambda_m R$ or, assuming that $\lambda_m \sim \frac{1}{\omega_{\mathbf{co}}}$, we get the needed correction (see also the discussion at the end of the next section). This should make it clear that our definition is a physical one: the outcome really depends on what particle is used to probe our screen. This will become even more important after we move to the discussion of possible corrections.

\subsection{Case of Weak Noncommutativity}

Now we pass to the calculation of the noncommutative corrections to the formula (\ref{areacommut}) for the case when noncommutativity is not too strong. To begin with, we would like to discuss the range of the applicability of our method. From the previous section, we already know that the cut-off scale, $\omega_{\mathbf{co}}$, should be much bigger then $1/R$ if we want to see any correction to the commutative result (otherwise any correction will be just masked by the experimental error). But there is another bound on $\omega_{\mathbf{co}}$ coming from the fact that this cut-off should still be well below $N/R$. This comes about due to the following reason: we still want that the fuzzy sphere, $S_F$, be not too fuzzy, i.e. $N$ still should be much bigger then 1. Then $N/R$ is just the order of the largest eigenvalue, see Eq.(\ref{spectrum}), and to be in the regime of corrections we need $\omega_{\mathbf{co}}$ to be much smaller then this largest eigenvalue. So, combining, we have the following range for the cut-off, where we expect to see corrections due to noncommutativity:
\begin{eqnarray}\label{range}
1\ll R\omega_{\mathbf{co}}\ll N \ .
\end{eqnarray}

Assuming that (\ref{range}) holds, we have the following leading correction to the classical result (\ref{classical})
\begin{equation}\label{classicalcorrected}
\Big(j_{max}+\frac{1}{2}\Big)^2= R^2 \omega_{\mathbf{co}}^2 \left( 1+\frac{R^2 \omega_{\mathbf{co}}^2}{N^2} + \mathcal{O}\left( \frac{1}{N^2},\frac{R^2 \omega_{\mathbf{co}}^2}{N^3} \right)\right)\ .
\end{equation}
Using this result in (\ref{counting2}), we arrive at the corrections to (\ref{areacommut}):
\begin{eqnarray}\label{areacorrected}
Area(S_F)\approx 4\pi R^2 \left( 1 + \frac{R^2 \omega_{\mathbf{co}}^2}{N^2}\right)\ .
\end{eqnarray}

Let us analyze the result (\ref{areacorrected}). First of all, we have to make sure that the noncommutative correction is seen on the background of the classical error, see (\ref{exact}) and the discussion after. This is equivalent to neglecting the second term in (\ref{counting2}). This means that while satisfying (\ref{range}), $\omega_{\mathbf{co}}$ should also respect the following
\begin{eqnarray}\label{range1}
\frac{R^2 \omega_{\mathbf{co}}^2}{N^2}\gg\frac{1}{R \omega_{\mathbf{co}}} \ .
\end{eqnarray}
Combined with (\ref{range}), this puts quite unrealistic restrictions on the possibility to observe these noncommutative corrections even in principle. To see this, we should answer the following question: what are $\omega_{\mathbf{co}}$ and $N$? We assume that $\omega_{\mathbf{co}}$ should be of the order of the inverse Compton wavelength of the test particle, which is used to probe the fuzzy sphere. This assumption seems very natural in view of the fact that this particle is the ``device'' used to probe the holographic screen in the entropic approach. As for $N$, we make another natural assumption that this is the same $N$ that is used in the formulation of the entropic scenario, i.e. the number of quanta of area, $N\sim A/l^2_P$. Now, if we require that the quantum corrections (\ref{areacorrected}) are seen on top of the classical ``experimental'' error (\ref{exact}), we immediately see that this is equivalent to
\begin{eqnarray}\label{smallcond}
\left(\frac{l_P}{\lambda_m}\right)^3 \gg\frac{R}{l_P} \ .
\end{eqnarray}
It is clear that (\ref{smallcond}) could be satisfied only in the regime of strong noncommutativity, which is well beyond of the assumed weak noncommutativity. So, we have to analyse the fully noncommutative model, i.e. Eq.(\ref{counting2}), without assuming (\ref{range}).

\subsection{Strong Noncommutativity}

\begin{figure}[htb]
\begin{center}
\leavevmode
\includegraphics[scale=0.7]{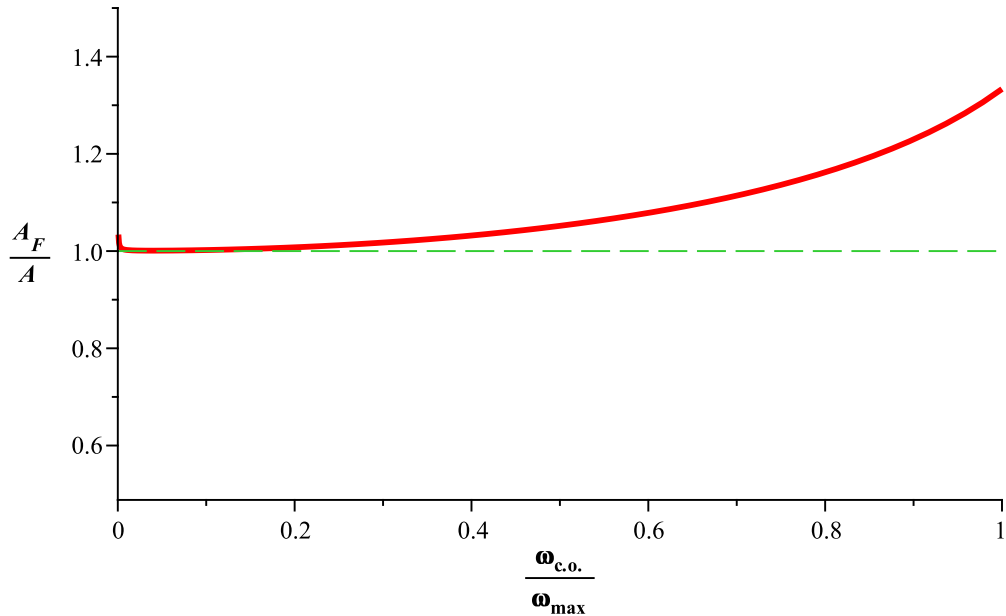}
\end{center}
\caption{The example of $A_F/A$ vs $\omega_{\mathbf{co}}$, where $A=4\pi R^2$ and $\omega_{max}$ is determined from (\ref{jmax}) and $N=100000$. }
\label{fig:scale1}
\end{figure}

Because in this regime it is hard to expect that our model will correctly describe quantum gravitational effects, we will just do some qualitative analysis.\footnote{This still should make sense because, as we mentioned before, noncommutativity is a nonperturbative residue of QG, so it should capture QG effects up to $\lambda_m \leq l_P$.} For this, let us plot the behaviour of the area as seen by the particle, $A_F$, versus the cut-off scale, $\omega_{\mathbf{co}}$. The result is shown on Fig.\ref{fig:scale1}. What does exactly this picture mean?

First of all, one should not be deceived by ``big enough'' $N$: $N=100000$ corresponds to highly noncommutative (i.e. quantum) regime. This is because for this $N$, the radius of the screen (using our assumption, $N\sim\frac{A}{l^2_P}$) is just three orders below the Plank scale.

Secondly, where is the Plank scale, $\omega_P\sim\frac{1}{l_P}$, on this figure? With the same assumptions as above, we can easily see that
\begin{eqnarray}\label{plankscale}
\frac{\omega_P}{\omega_{max}}\sim \frac{1}{\sqrt{\omega_{max}R}}\sim \frac{1}{\sqrt{N}} \ .
\end{eqnarray}
So, even for such a highly quantum regime $\frac{\omega_P}{\omega_{max}}\sim 10^{-3}$, i.e. the Plank scale is well below $\omega_{max}$.

\begin{figure}[htb]
\begin{center}
\leavevmode
\includegraphics[scale=0.7]{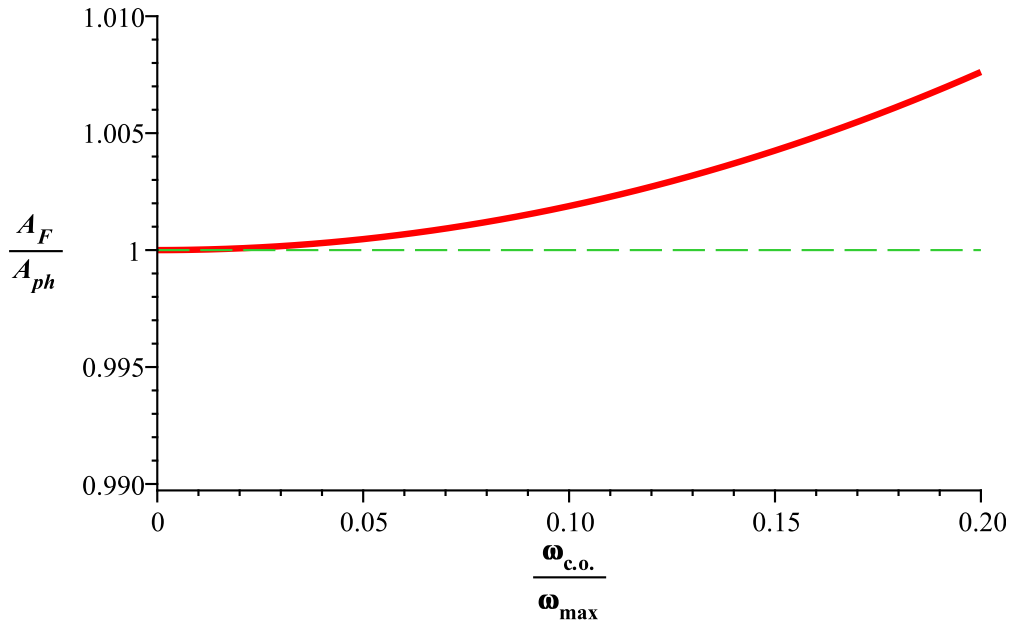}
\end{center}
\caption{Noncommutative area (normalized to $A_{ph}$ - the classical ``physical'' area, which includes the ``experimental'' error (\ref{exact})) as a function of the cut-off scale. Significant deviation is seen well below the cut-off.}
\label{fig:scale2}
\end{figure}

Thirdly, we can see on Fig.\ref{fig:scale2} that deviations from the classical area (plus experimental error) defined by (\ref{exact}) start at quite high cut-off, which in this specific case is well above the Planck scale. This is in complete agreement with (\ref{range}), which could be re-written as $\frac{1}{N}\ll \frac{\omega_{\mathbf{co}}}{\omega_{max}} \ll 1$, and the conclusion at the end of the previous section that for this range, noncommutative corrections are not seen on the background of the experimental uncertainty. So, we can say that even if a test particle could probe the Plank scale\footnote{One can imagine using as a test particle the so called maximon \cite{Markov:1987yu}, i.e. a speculated elementary particle such that its Compton wavelength equals to its gravitational radius.} it will see almost classical area (for this value of $N$). We can see that due to (\ref{plankscale}), the situation will drastically change if one goes deeper into quantum regime, i.e. smaller $N$. (See the discussion in the next section.)

All the above seems to indicate that there are no significant corrections to the physical (i.e. as seen by a test particle) area of the holographic screen due to noncommutativity of this screen. As the result, one might conclude that the only corrections to Newton's law in entropic scenario are due to the corrections to entropy as in (\ref{LQGentropy}-\ref{Newtoncorrected2}). In the next section we will discuss wether it is true or not.

\section{Discussion and Interpretation}\label{Discussion}

Let us analyze what we have obtained. We will start by confronting our results with the result (\ref{Newtoncorrected2}), which was also obtained in the noncommutative framework. Looking at (\ref{Newtoncorrected2}), we immediately notice that the only way to see any sizable corrections is to approach the source mass $M$ by distance of order of the Plank scale, otherwise any such correction will be exponentially suppressed. What is this regime in our picture? It is not hard to see that it corresponds to $N\sim 1$. But, according to (\ref{plankscale}), this is exactly where the maximal cut-off scale $\omega_{max}$ is of order of the Plank scale! Then it is pretty obvious that the quantum corrections to area will be very significant. E.g., if $N=100$, the corrections in (\ref{Newtoncorrected2}) will be suppressed by the factor of the order of $e^{-100}$, while looking at Fig.\ref{fig:scale2} (which looks pretty much the same for $N=100$, but now the Plank scale is around $0.2\omega_{max}$) we see that noncommutative corrections to area will be of order of 1\%. In this regard, it is important to understand that to see these corrections one needs a test particle that can probe the Plank scale. But while (\ref{Newtoncorrected2}) produces corrections only within the Plank distance of the source, the noncommutative corrections to area are non-zero even away from the origin. This leads to the conclusion that these noncommutative corrections can completely shadow the effects due to the corrections to entropy (given by (\ref{Newtoncorrected2})).

Let us now look at (\ref{Newtoncorrected1}). This type of corrections looks much more reliable. Namely, let us consider the first correction (due to the logarithmic term in (\ref{LQGentropy})). It is pretty obvious from our analysis that there exists some range where this term will clearly dominate any correction to area. This is because this term behaves as $1/N$ and does not depend on the Compton wavelength of a test particle. So, when the test particle has the large Compton wavelength compared to the Plank scale (but still small enough to probe such distances, i.e. $\lambda_m\ll R$, see (\ref{range}) and the discussion after (\ref{exact})), one will have sizable corrections to Newton's law while almost negligible corrections to area (see Fig.\ref{fig:scale2} and Eq.(\ref{plankscale})). But still, closer to the Plank scale this will be completely masked by the area correction. This could be interpreted in the following way: it is well known that the same corrections come also from perturbative quantum gravity \cite{Donoghue:1994dn,Hamber:1995cq}, so it is reasonable to believe that they should be trusted well above the Plank scale. Closer to the Plank scale perturbative calculations would clearly fail and this is where noncommutative effects, which are nonperturbative traces of QG, will start to matter.\footnote{It is worth remembering that (\ref{exact}) was calculated for a very specific choice of the Dirac operator. As the result, the absence of $1/N$ corrections in (\ref{areacorrected}) is very much accidental. Other choices of the Dirac operator might easily produce these corrections leading to a much stronger deviation from the classical area. In this case, even the term in Newton's law due to the logarithmic correction could be overshadowed by the area corrections.}

This consideration brings our attention to the very important point - the non-trivial dependence of the possible corrections on a test particle. A test particle now becomes the essential part of the definition of gravity. It is needed not just to reveal already existing gravity (in the form of curvature of space-time as in GR), but to ``produce'' it by changing entropy of the holographic screen, which leads to the entropic force. By looking at the holographic screen from the point of view of noncommutative geometry, it makes this special role of a test particle even more obvious: now, the same holographic screen will look differently for different test particles. This looks quite like the violation of the equivalence principle by the quantum gravitational effects. In our approach, this is reflected in the result that the area corrections depend on the physical cut-off, which is a function of the mass of a test particle, $\omega_{\mathbf{co}}\sim m$.

Before we close this section, we would like to support the assumption made at the very beginning: the cut-off scale is always below the ``end'' of the spectrum. Using considerations as in \cite{Dvali:2010jz}, one can argue that the cut-off scale should be less or equal to the Plank scale. But we have seen that the Plank scale is always below the maximal eigenvalue (at least in our model). Thus, from the point of view of any experiment (i.e. from the operational point of view) one could never tell whether the spectrum is finite or not. What one can only do is to measure possible deviations in geometrical quantities based on the deviation of the \textit{observed} part of the spectrum from the classical one.

\section{Summary and Conclusions}

In this paper we analyzed the possible effects of noncommutativity in the entropic scenario by using a fuzzy sphere as a holographic screen. In contrast with the other efforts in this direction, which deal with the corrections to entropy (and, as the consequence, to the apparent gravitational force), we concentrate our attention on the question of the interaction of a test particle and a holographic screen. That this is very important follows from the special role played by a test particle in the entopic scenario, which is rather different from its role in GR, based on the equivalence principle.\footnote{To demonstrate a very special role of a test particle in this scenario as well as the importance of the knowledge of the microscopic dynamics of the interaction of the particle with a screen, we could imagine the following Gedankenexperiment: let us consider a test particle with some Compton wavelength $\lambda_m$. Typically, there will be several holographic screens on this length (in our case the number of the screens can be estimated as $\frac{\lambda_m R}{l^2_P}$, which is huge away from the Plank scale). Then the question is: to the entropy of which of those screens does this test particle contribute? We can answer this question only if we know the details of the microscopic dynamics of the screen-particle system.}

In the absence of the necessary apparatus to directly study the process during which a test particle ``becomes a part of the screen'', we make an  attempt to study how this screen is seen by the particle. For this we adopt the model of a noncommutative screen that, as we argued, should capture some nonperturbative QG effects. As the main tool, we use the generalization of Weyl's theorem, which has proven to be quite efficient in the study of deformed geometries.

The main conclusion of this paper could be formulated as follows:

While perturbative corrections (as the second term in (\ref{Newtoncorrected1})) can be trusted well below the Plank scale (in conformity with their universal model independence), neither correction should be trusted close to the Plank scale. In particular, it seems, that the corrections given by (\ref{Newtoncorrected2}) will be washed away by the uncertainties due to our ignorance about the details of the interaction between a screen and a test particle.

Our discussion opens up the way to the speculation about the possible violations of the equivalence principle. But this could happen only on the Plank scale, so it is hard to believe in any possibility of the experimental confrontation.

To conclude, our study showed that no effects of noncommutativity (which encodes at least some nonperturbative QG effects) will be seen below the Plank scale, but when one approaches this scale they will start to dominate over the perturbative corrections. To get further control over these effects, one needs to use the full quantum gravity in the form of strings, loops or any other.

\section*{Acknowledgements}

\hspace{0.5cm}AP acknowledges partial support of CNPq under grant no.308911/2009-1.
CMG is supported by CNPq master's scholarship.

 \end{document}